\documentclass[3p,times,procedia]{elsarticle}
\usepackage{nupha_ecrc}

\volume{00}

\firstpage{1}

\journalname{Nuclear Physics A}

\runauth{
}

\jid{nupha}

\jnltitlelogo{Nuclear Physics A}

\usepackage{amssymb}

\usepackage[figuresright]{rotating}

\begin{document}

\begin{frontmatter}



\dochead{}

\title{
In search of chiral magnetic effect: separating flow-driven background effects and quantifying anomaly-induced charge separations 
}


 \author[1]{Xu-Guang Huang\footnote{Presenter}}
 \address[1]{Physics Department and Center for Particle Physics and Field Theory, Fudan University, Shanghai 200433, China.}

 \author[2]{Yi Yin}
\address[2]{Physics Department, Brookhaven National Laboratory, Upton, NY 11973, USA}

 \author[3,4]{Jinfeng Liao}
\address[3]{Physics Department and Center for Exploration of Energy and Matter,
Indiana University, 2401 N Milo B. Sampson Lane, Bloomington, IN 47408, USA.}
 \address[4]{RIKEN BNL Research Center, Bldg. 510A, Brookhaven National Laboratory, Upton, NY 11973, USA}


\begin{abstract}
We report our recent progress on the search of Chiral Magnetic Effect (CME) by developing new measurements as well as by hydrodynamic simulations of CME and background effects, with  both approaches addressing the pressing issue of separating flow-driven background contributions and possible CME signal in current heavy ion collision measurements.
\end{abstract}


\end{frontmatter}


\section{Introduction}
\label{sec:intro}
Magnetic field $\bf{B}$ of extreme strength is created in non-central heavy-ion collisions by fast moving nuclei.
Its magnitude is on the order of $eB\sim m^{2}_{\pi}$ and points approximately in the out-of-plane direction.
The phenomenological consequences of such unprecedented    magnetic field have attracted much recent interest (see e.g.~\cite{Kharzeev:2015znc,Huang:2015oca,Liao:2014ava} for recent reviews).
One particularly important example is the chiral magnetic effect (CME) -- the generation of charge current ${\bf J}$ along magnetic field ${\bf B} $ for a chiral medium in the presence of axial charge imbalance: ${\bf J} = C_A  \mu_A {\bf B}$ where $\mu_A$ characterizes chirality imbalance.
 The possible occurrence of axial charge imbalance is related to a salient feature of QCD as  non-Abelian gauge theory, in which   topologically nontrivial gauge field configurations such as instantons and sphalerons are known to exist and play curial roles for nonperturbative dynamics.
When coupled to massless fermions (e.g. light quarks in QCD), the fluctuations of topological charges carried by those configurations will translate into fluctuations of chirality imbalance via chiral anomaly. Remarkably, the coefficient $C_A$ in the above CME relation is completely fixed from anomaly relation.
Therefore CME opens a unique possibility of observable effects that may manifest  nontrivial topology and chiral anomaly of QCD in heavy-ion collisions.

The CME will contribute to the reaction-plane dependent azimuthal correlation observables:
\begin{eqnarray}
 \label{gamma_ob}
\gamma_{\alpha\beta}= \langle\cos(\phi_i+\phi_j-2\Psi_{\rm RP})\rangle_{\alpha\beta}\, , \qquad
\delta_{\alpha\beta}= \langle\cos(\phi_i-\phi_j)\rangle_{\alpha\beta}\, ,
\end{eqnarray}
with $\alpha,\beta=\pm$ labeling the  species and $\phi_{i,j}$  the azimuthal angles of two final state charged hadrons. The $\Psi_{\rm RP}$ denotes reaction plane angle and we set $\Psi_{\rm RP}=0$.
However, these observables also suffer from   elliptic flow driven background contributions and can not be entirely attributed to CME.  An outstanding challenge at present is to quantitatively decipher possible CME signals from the measured correlation observables.
In this contribution we will first discuss a recent proposal of using the Cu + Au collisions to test the relative importance between electromagnetic-field contributions and the flow-driven contributions. We will then report our recent efforts to   quantify the influence of the key theoretical uncertainties (like initial axial charge fluctuations and magnetic field lifetime) on CME signals and to   further separate a major background effect (transverse momentum conservation effects) in measured correlations along with predictions for future measurements, by using an anomalous and viscous hydrodynamic simulation framework.

\section{CME in Cu + Au collisions and charge-dependent correlations}
\label{sec:Cu-Au}
One of the difficulties in separating the CME and the elliptic-flow driven contributions in Au + Au collisions or Pb + Pb collisions is that in such collisions the strength of the magnetic fields have a similar dependence on centrality as  the elliptic flow. It was proposed to use the U + U collisions to overcome this difficulty because in the most central U + U collisions the magnetic fields almost die away while the elliptic flow is still sizable~\cite{Voloshin:2010ut,Wang:2012qs,Bloczynski:2013mca}. We here report another proposal that the Cu + Au collisions may also provide insight to disentangling the CME and the elliptic-flow-driven contributions.

\begin{figure}[hbt!]
\begin{center}
\includegraphics[height=0.2\textwidth]{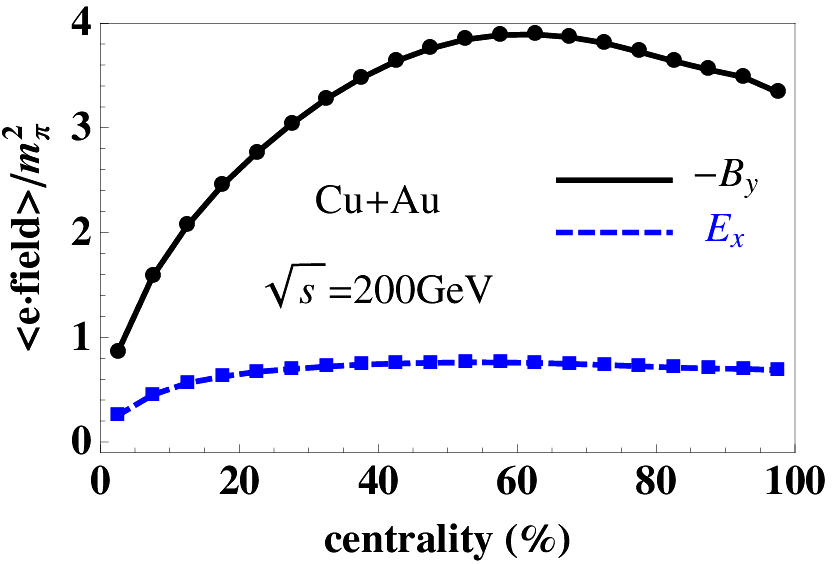}$\;\;\;\;\;\;\;\;\;\;\;\;\;\;\;\;\;\;\;\;\;\;$
\includegraphics[height=0.2\textwidth]{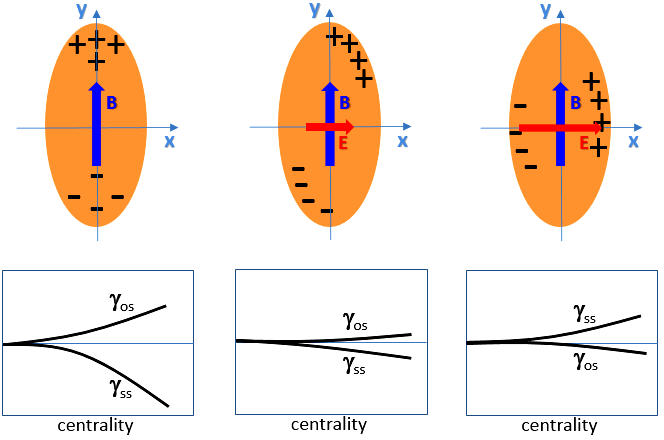}
\caption{(Left) The electric and magnetic fields in Cu + Au collisions. (Right) Illustration of the charge dipole induced by CME and electric field in Cu + Au collisions. 
\label{illu}
}  \vspace{-1cm}
\end{center}
\end{figure}

Here, the main ingredient of Cu + Au collisions is the in-plane electric fields pointing from the Au to Cu nuclei; see Fig.~\ref{illu} (Left) for the numerical result~\cite{Deng:2014uja}. Suppose $\gamma_{\alpha\beta}$ is dominated by CME rather than the elliptic-flow driven effects. Then in non-central Cu + Au collisions, the in-plane electric fields will induce an in-plane charge separation in addition to the out-of-plane charge separation due to CME; see Fig.~\ref{illu} (Right) for illustration. The presence of the in-plane dipole will suppress the strength of $\gamma_{\alpha\beta}$; and if the electric fields are strong enough, or somehow equivalently, if the lifetime of the electric fields is long enough, the total charge dipole will be mainly along the in-plane direction and the signs of the same-sign and opposite-sign correlations will be reversed. On the other hand, if $\gamma_{\alpha\beta}$ is dominated by the elliptic-flow driven effects, we expect that $\gamma_{\alpha\beta}$ (more precisely, $\Delta\gamma\equiv\gamma_{\rm OS}-\gamma_{\rm SS}$ with $\gamma_{\rm OS/SS}$ the opposite-sign/same-sign correlation, because the directed flow in Cu + Au collisions may contribute to both $\gamma_{\rm SS}$ and $\gamma_{\rm OS}$.) would not change too much from Au + Au collisions to Cu + Au collisions. Plausibly, $\Delta\gamma$ as a function of centrality in Cu + Au collisions would lie between that in Cu + Cu and Au + Au collisions. Thus the Cu + Au collisions may provide a way to test the relative importance of the electromagnetic fields and the elliptic flow in the   correlations $\gamma_{\alpha\beta}$. More information is given in ~\cite{Deng:2014uja}.

\section{Toward quantitative understanding of the same-charge correlation}
\label{sec:same-sign}
We now turn to discuss our recent progress on separating background contributions from the data and quantitatively computing CME signal~\cite{Yin:2015fca}.
Charge separation along magnetic field induced by the CME would contribute to the azimuthal distribution of charged hadrons as follows:
\begin{eqnarray}
\left[ {dN^{H}}/ {d\phi} \right]_{\rm CME}
\propto
[1+ 2 Q^H a^{H}_{1}\sin(\phi) +...]\, .
\end{eqnarray}
Here, ``H''
labels the species of the hadron,
e.g. $H=\pi^{\pm}, K^{\pm},...$.
Our first task is to \textit{quantitatively} relate the magnetic field $eB$  and initial axial charge density $\mu_{A}$ in the early stage of heavy-ion collisions to the  $a^{H}_{1}$ in the late stage hadron distribution.
To that end,
we have solved anomalous hydrodynamics for axial and vector charge density evolution on top of background hydrodynamic profile  from data-validated VISH code~\cite{hydro}.

The computed results for $a^{H}_{1}$ depend on two inputs, a) initial axial charge density $n_{A}$ and b) the magnetic field strength and lifetime.
We take initial axial charge density $n_{A}$ to be proportional to initial entropy $s_{I}$.
We also take $B$ to be homogeneous in the transverse plane and introduce $\tau_{B}$ that controls its lifetime:
\begin{equation}
\lambda_{A} \equiv  n_{A} / (N_{f}s_{I}) \approx  Q_{A} / (N_{f}S) \, ,
\qquad
eB(\tau)= (eB_0)/ [1+\left(\tau/\tau_{B}\right)^{2}]\, .
\end{equation}
Here $Q_{A}$ is the total initial axial charge and $S$ the total entropy.
The peak value $eB_{0}$  is taken from Ref.~\cite{Bloczynski:2012en}.

\begin{figure}[hbt!]
\includegraphics[width=0.31\textwidth,height=0.2\textwidth]{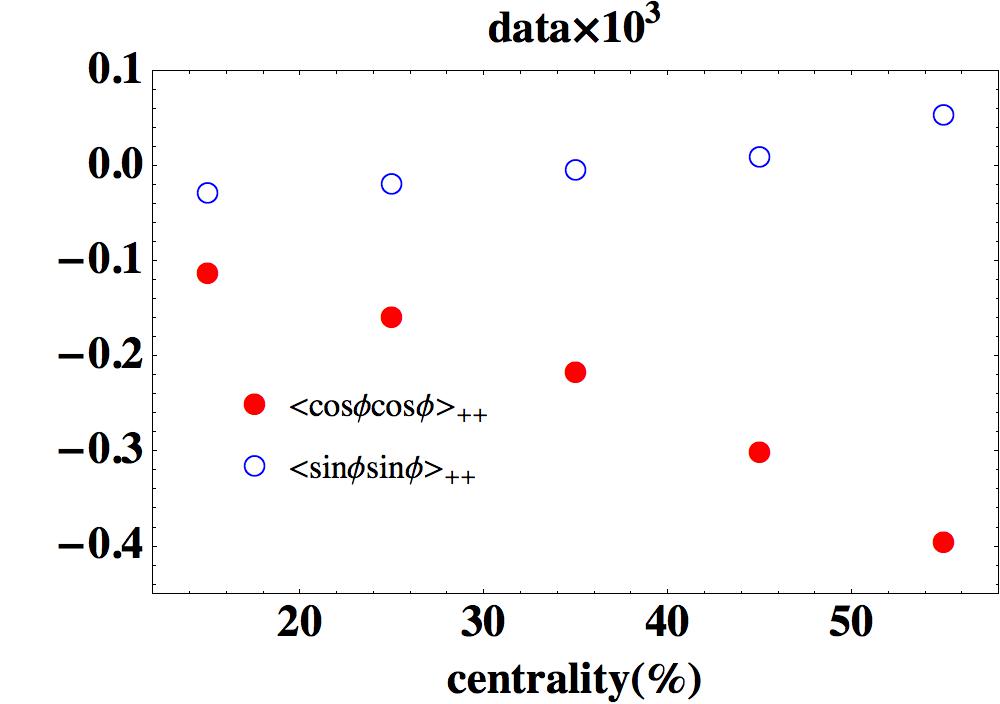}
\includegraphics[width=0.31\textwidth,height=0.2\textwidth]{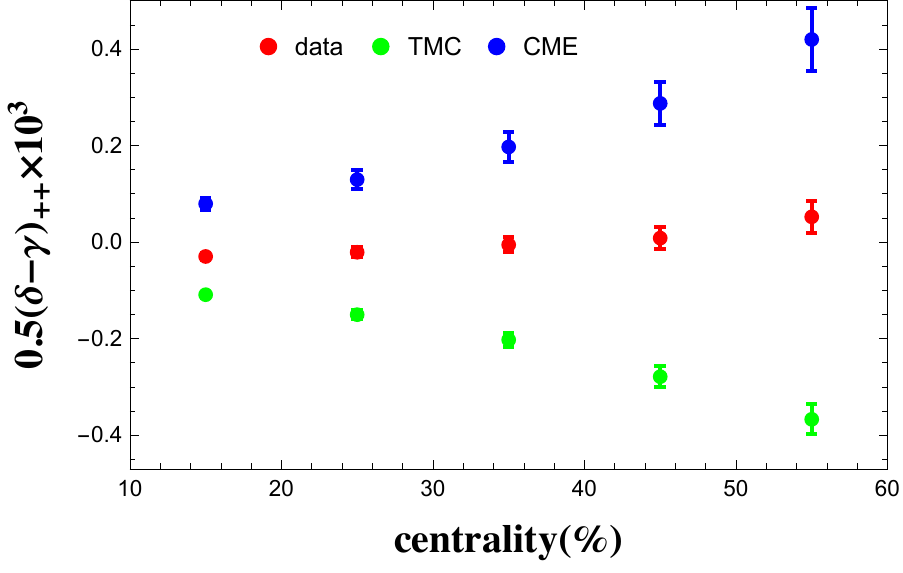}
\includegraphics[width=0.31\textwidth,height=0.2\textwidth]{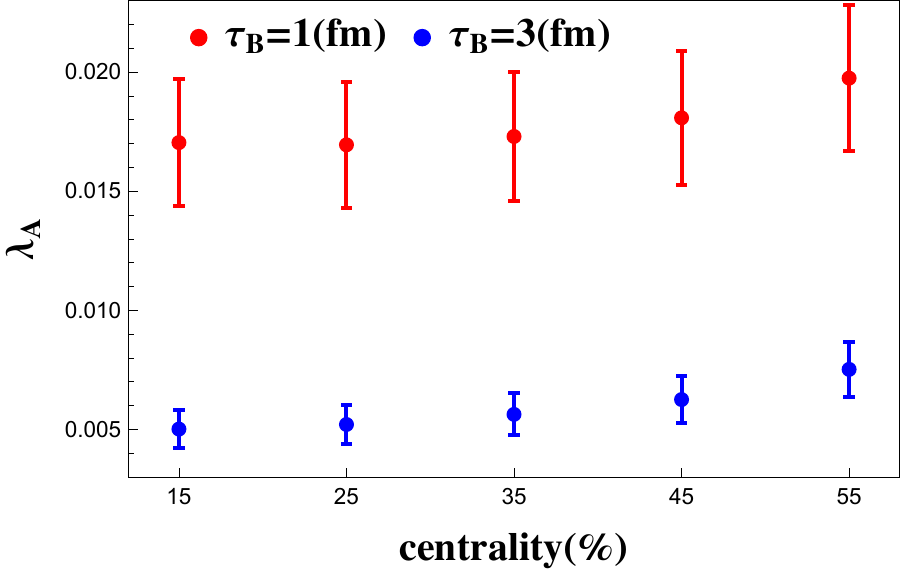}
\caption{(Color online)
(Left) STAR data for same sign charge correlations.
(Middle)
Same sign $(\delta-\gamma)/2$ data and its decomposition into CME and TMC contributions for different centrality.
(Right) Centrality dependence of extracted initial axial charge parameter $\lambda_{A}$ for different values of  $\tau_{B}$.
All error bars originate from the uncertainty in the STAR data.
\vspace{-0.05in}
\label{fig_data}
}
\end{figure}

How would $a_{1}$ depend on $\lambda_{A}$, $eB_0$ and $\tau_{B}$?
We have found from our numerical computations that $a_{1}$ is approximately proportional to $\lambda$ and $eB$. This may not be surprising given the CME relation ${\bf J}\propto {\mu_A} {\bf B}$.   Furthermore, our results suggest that $a_{1}$ grows linearly when $\tau_{B}$ is small and becomes saturated when $\tau_{B}$ gets comparable to the life time of the fireball. See Ref.~\cite{Yin:2015fca} for more details.

Next let us examine the background contributions to the measured correlations.
This is an essential step towards extracting CME signal from the data.
We focus on the reaction-plane-projected same-charge pair correlations,  obtained via   $(\gamma+\delta) = \langle\cos\phi\cos\phi\rangle$ and
$(\gamma-\delta) =  \langle\sin\phi\sin\phi\rangle$.
Those data measured by STAR are shown in Fig.~\ref{fig_data} (left).
 The CME is expected  to mainly contribute to $\langle\sin\phi\sin\phi\rangle_{\rm CME} \simeq a^{2}_{1}$ and hence $\langle\cos\phi\cos\phi\rangle_{\rm CME}\simeq 0 $.
In other words,
without any background effect,  $\langle\sin\phi\sin\phi\rangle>0,\langle\cos\phi\cos\phi\rangle\approx 0$.
In contrast,
$\langle\cos\phi\cos\phi\rangle$ is negative and $\langle\sin\phi\sin\phi\rangle$ is around $0$ from the data.
This implies the existence of a negative background contribution to same-charge correlation.
and motivates a possible interpretation of the data:
$
\langle\cos\phi\cos\phi\rangle \approx\textrm{[negative background]}\,\, ,
\langle\sin\phi\sin\phi\rangle \approx\textrm{[negative background}+\textrm{positive CME}]\,
$.

Past studies have suggested that a major background contribution to the same sign correlation is the transverse momentum conservation (TMC) effect~\cite{Bzdak:2010fd}.
We generalize analytic formula for single-component TMC in \cite{Bzdak:2010fd} to the case of multiple types  of hadrons.
The results depend on the elliptic flow of the hadrons which can be directly computed from the {\em same} hydrodynamic background we use for CME computation.
It further depends on an overall parameter $N_{\rm TMC}$ (the total number of {\em all produced}, rather than just measured, particles)  which we  take  as an input parameter that controls the magnitude of TMC effect.
By fitting $\langle\cos\phi\cos\phi\rangle$  from the data,
we fix $N_{\rm TMC}$.
We next compute TMC contribution to $\langle\sin\phi\sin\phi\rangle$ with already fixed  $N_{\rm TMC}$ as inputs.
Subtracting them from the data, we finally obtain CME signals.
Those two contributions are summarized in Fig.~\ref{fig_data} (middle).
The above procedure is also schematically sketched below:
\begin{eqnarray}
\gamma_{\rm data}+\delta_{\rm data} &\approx& \gamma_{\rm TMC}+\delta_{\rm TMC}\,\,\,\, \stackrel{\textrm{fitting}}{\Longrightarrow}
\textrm{TMC parameters}\,\,\,\,\Longrightarrow \,\,\,\,  \gamma_{\rm TMC}-\delta_{\rm TMC}
\nonumber \\
&\stackrel{\textrm{subtraction
 from data} }{\Longrightarrow} & \gamma_{\rm CME}-\delta_{\rm CME}\propto a^{2}_{1}\, ,
\end{eqnarray}

With the exacted CME signal from data   we now could quantify the initial axial charge parameter  $\lambda_{A}$ by comparing  the signal with  anomalous hydrodynamics results.
In Fig.~\ref{fig_data} (right),
we plot $\lambda_{A}$ at different centralities with two representative $\tau_{B}=1, 3$~fm.
They correspond to two different scenarios: the lifetime of magnetic field is very short and the life time of magnetic field is comparable to the fireball, respectively.
Few remarks are in order.
First, a shorter lifetime of magnetic field can be compensated by a larger initial axial charge density.
Second, the order of magnitude of $\langle Q^{2}_{A}\rangle$ is inline with the theoretical estimates based on Chern-Simons diffusion rate for gluonic topological fluctuations.
Finally, $Q^{2}_{A}$ increases mildly from central toward peripheral collisions: a smaller system typically has a larger fluctuations.
Our results also suggest that CME signal indeed can be used to extract the information on topological fluctuations.

It may be interesting to compare the two recent anomalous hydrodynamic studies for CME: \cite{Hirono:2014oda} (HHK) and \cite{Yin:2015fca} (YL). While HHK solves full evolution for charge densities (albeit with not-so-realistic bulk evolution and EoS), YL solves the linearized evolution (albeit on top of a data-validated bulk evolution). While HHK emphasizes event-by-event   simulations with glasma initial conditions, YL emphasizes simultaneous computation  of backgrounds which allows extraction of CME signal and constraints on initial conditions. Both studies, together in a complementary way, have made crucial steps for quantitative CME simulations.

\begin{figure}[hbt!]
\label{fig:pred}
\begin{center}
\includegraphics[width=0.3\textwidth,height=0.2\textwidth]{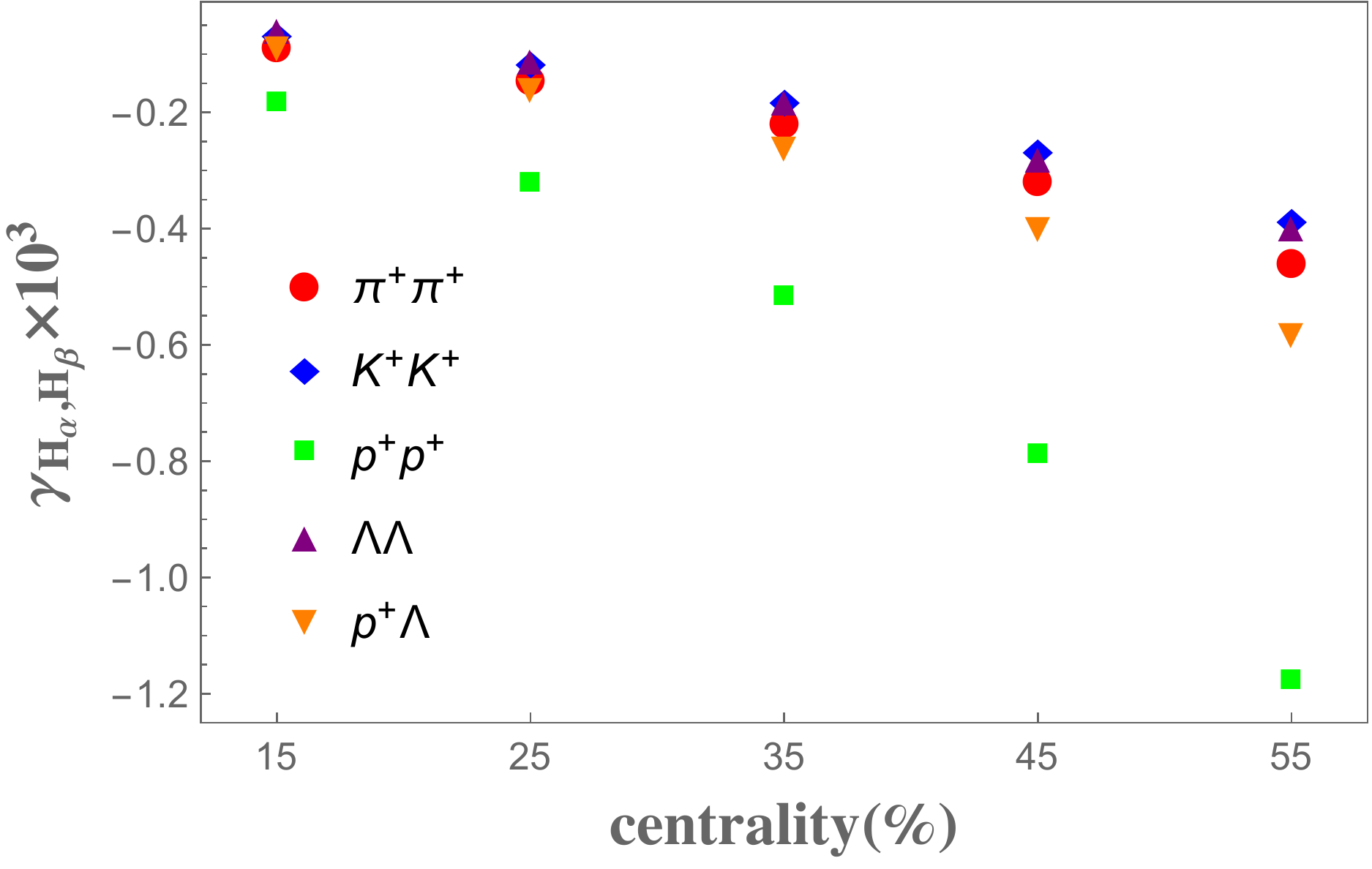}$\;\;\;\;\;\;\;\;\;\;$
\includegraphics[width=0.3\textwidth,height=0.2\textwidth]{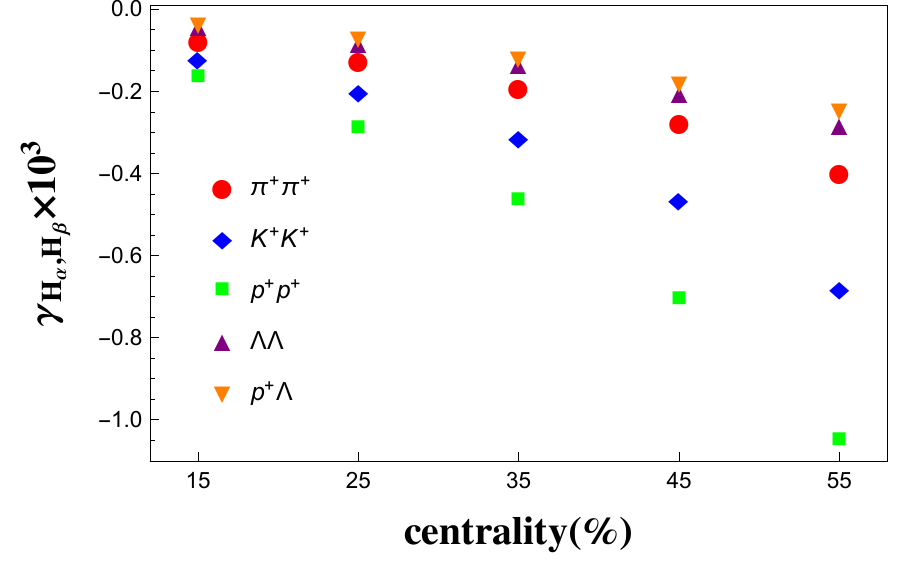}
\caption{(Color online)
Predictions for correlation $\gamma_{H_{\alpha},H_{\beta}}$ vs centrality. (Left) Two flavor (Rigt)Three flavor. See text for details.
}\vspace{-0.2in}
\end{center}
\end{figure}

Finally, we propose to use the same-charge azimuthal correlations for various identified hadron species as a nontrivial further test.
With our model parameters already fixed above, we can make quantitative predictions to be confronted by future measurements.
We have done computations for both the two-flavor case  and the three-flavor case (where the strange quarks also contribute to CME), with the results  summarized in Fig.~\ref{fig:pred}.
We note that   certain  observables (like  $\gamma_{K^{+}K^{+}}$) are very sensitive to potential strangeness contributions.

\section{Summary}

In summary, the most pressing issue in current search of the Chiral Magnetic Effect in heavy ion collisions, is the  separation of flow-driven background contributions and possible CME signal in the measured azimuthal correlations. We have reported our recent progress on addressing this issue by proposing new measurements that would be sensitive to flow-driven versus magnetic-field-driven effects as well as by  quantitatively computing both types of contributions for comparison with data.

\section*{Acknowledgements}
This work was supported by in part by DOE Grant No. DE- AC02-98CH10886 (YY), by NSF Grant No. PHY-1352368 as well as RIKEN BNL Research Center (JL), and by Shanghai Natural Science Foundation Grant No. 14ZR1403000 as well as the Young 1000 Talents Program of China (XGH).





\bibliographystyle{elsarticle-num}



\end{document}